\documentclass{emulateapj}
\usepackage{xspace}
\usepackage{amsmath}
\usepackage{hyperref}
\usepackage{framed} 
\usepackage{txfonts}
\usepackage{epstopdf} 

\slugcomment{To be submitted to the Astrophysical Journal}

\usepackage{rotating}
\usepackage{natbib}

\newcommand{\kmsmpc}{\>{\rm km}\,{\rm s}^{-1}\,{\rm Mpc}^{-1}}

\newcommand{\mpch}{\>h^{-1}{\rm {Mpc}}}

\newcommand{\beq}{\begin{equation}}
\newcommand{\eeq}{\end{equation}}

\newcommand{\g}{{\rm g}}

\newcommand{\avg}[1]{\langle #1 \rangle}

\newcommand{\drm}{{\rm d}}

\newcommand{\avn}{\avg{N}_M}

\newcommand{\tofour}{{\rm T04}}

\newcommand{\true}{{\rm true}}
\newcommand{\logh}{\log h}
\newcommand{\ft}{\rm fit}

\def\gtsima{$\; \buildrel > \over \sim \;$}
\def\ltsima{$\; \buildrel < \over \sim \;$}
\def\prosima{$\; \buildrel \propto \over \sim \;$}
\def\gsim{\lower.7ex\hbox{\gtsima}}
\def\lsim{\lower.7ex\hbox{\ltsima}}
\def\simgt{\lower.7ex\hbox{\gtsima}}
\def\simlt{\lower.7ex\hbox{\ltsima}}
\def\simpr{\lower.7ex\hbox{\prosima}}
\def\la{\lsim}
\def\ga{\gsim}
\def\lta{\la}
\def\gta{\ga}

\newdimen\hssize
\newdimen\hdsize
\shorttitle{Galaxy bias from SDSS}
\shortauthors{More S.}

\begin{document}

\title{How accurate is our knowledge of the galaxy bias?}

\author{Surhud \ More\altaffilmark{1,2}}

 \altaffiltext{1}{Kavli Institute for Cosmological Physics and Enrico
 Fermi Institute, The University of Chicago, 5640 S. Ellis Ave.,
 Chicago-60637, IL, USA;}
 \altaffiltext{2}{surhud@kicp.uchicago.edu}

\begin{abstract}
Observations of the clustering of galaxies can provide useful
information about the distribution of dark matter in the Universe. In
order to extract accurate cosmological parameters from galaxy surveys,
it is important to understand how the distribution of galaxies is
biased with respect to the matter distribution. The large-scale bias of
galaxies can be quantified either by directly measuring the
large-scale ($\lambda\gta60~\mpch$) power spectrum of galaxies or by
modeling the halo occupation distribution of galaxies using their
clustering on small scales ($\lambda\lta30~\mpch$). We compare the
luminosity dependence of the galaxy bias (both the shape and the
normalization) obtained by these methods and check for consistency.
Our comparison reveals that the bias of galaxies obtained by the small
scale clustering measurements is systematically larger than that
obtained from the large scale power spectrum methods. We also find
systematic discrepancies in the shape of the galaxy bias-luminosity
relation. We comment on the origin and possible consequences of these
discrepancies which had remained unnoticed thus far.
\end{abstract}



%
\section{Introduction}
\label{sec:intro}

Large scale galaxy redshift surveys provide a unique window to probe
the distribution of dark matter. The abundance and clustering of
galaxies contain enormous information about the dark matter
distribution in the Universe and hence various cosmological parameters
\citep[e.g., see][]{vdb2003,vdb2007,Tinker2005,Cacciato2009}. However,
in order to use galaxies as cosmological probes, an accurate knowledge
of how the galaxy distribution is biased with respect to the dark
matter distribution is essential.

The Sloan Digital Sky Survey \citep[][SDSS hereafter]{York2000} has
imaged nearly a quarter of the sky ($\sim$ 8000 sq. deg.) and obtained
accurate spectroscopic redshifts for more than 0.9 million galaxies
over a span of 8 years of its operation \citep{Abazajian2009}. The
availability of the three dimensional positional information for such
a large sample of galaxies has enabled an accurate measurement of the
galaxy power spectrum over more than an order of magnitude in
wavelength range \citep[][T04 hereafter]{Tegmark2004}. The shape of
the power spectrum of galaxies has been used in combination with the
results from the cosmic microwave background to put stringent
constraints on cosmological parameters such as the matter density
parameter, $\Omega_m$, the vacuum energy density parameter,
$\Omega_\Lambda$, its equation of state, $w_0$, and constraints on
quantities relevant to particle physics such as the sum of masses of
neutrinos \citep[e.g., see][]{Tegmark2004b,Seljak2005,Reid2010}.

The cosmological parameter $\sigma_8$ describes the variance of the
matter density distribution smoothed on a scale of $8~\mpch$
(comoving) at an early time in the Universe extrapolated to today
using linear perturbation theory. The quantity $\sigma_8^2$ is
proportional to the amplitude of the matter power spectrum. As
galaxies are biased tracers of the matter distribution, observations
of galaxy power spectra on large scales ($\lambda\gta60~\mpch$) can
only constrain the product of galaxy bias and the parameter
$\sigma_8$.

This degeneracy between $\sigma_8$ and galaxy bias can be broken by
information from the clustering of galaxies (or by the clustering of
the dark matter around galaxies probed by galaxy-galaxy lensing) on
small scales ($\lambda\lta30~\mpch$). However the clustering on these
scales is inherently difficult to model. The halo model framework has
been routinely used to carry out such modeling \citep[e.g.,
see][]{Cooray2002}. The halo occupation distribution (HOD) of galaxies
constrained by these models using the small-scale clustering predicts
the large-scale bias of galaxies whose product with $\sigma_8$ should
agree with the power spectrum measurements modulo the systematic
uncertainties in either of the methods.

In this paper, we compile and critically compare the measurements of
the luminosity dependence of galaxy bias on large scales obtained
using the measurements of the galaxy power spectrum and those obtained
by modeling the clustering of galaxies on small scales. This allows
us to gauge the systematic differences in the galaxy bias obtained
from these different methods. In Section \ref{sec:diff}, we briefly
describe the different methods used to obtain galaxy bias and compare
the results after accounting for the differences in the cosmological
parameters used by each of the measurements. In
Section~\ref{sec:discuss}, we discuss the various sources of
systematics in these methods and summarize our findings.

\section{Measurements of Galaxy bias}
\label{sec:diff}

\subsection{Galaxy power-spectrum measurements}
\label{sec:diff_dr2}

T04 presented the measurement of the power spectrum of galaxies from
the entire (flux-limited) main galaxy sample from SDSS-DR2
\citep{Abazajian2004}. They employed a matrix-based method which uses
the pseudo-Karhunen-Lo\`{e}ve eigenmodes to produce a minimum variance
measurement of the power spectrum of galaxies. The authors corrected
the galaxy power spectrum measurements for redshift-space distortions
and for the artificial red-tilt of the galaxy power spectrum caused by
the luminosity dependence of bias, an effect which we discuss below.

The clustering of galaxies has been observed to depend upon different
galaxy properties such as their luminosity and color. The luminosity
dependence of clustering can cause a scale dependence in the
power spectrum even on large scales ($\lambda\gta60~\mpch$). For a
flux-limited survey such as the SDSS, the measurement of the galaxy
power spectrum on large scales is dominated by bright
galaxies which are observed to larger distances whereas the galaxy
power spectrum on smaller scales is dominated by the dim ones. As the
clustering of galaxies increases with their luminosity, this effect
can cause the measured power spectrum to be redder than the true power
spectrum. To correct for the above effect, T04 measured the luminosity
dependence of the galaxy bias using volume limited samples of galaxies
binned by luminosity.  Specifically, they measured the bias of
galaxies of a given luminosity $L$ divided by the bias of $L_*$
galaxies and provided the following fitting function to summarize
their result 
\begin{equation}
\label{eq:teg04}
\frac{b}{b_*} = A + B\frac{L}{L_*^{\tofour}} +
C\,(^{0.1}M_r-M_*^{\tofour})\,.
\end{equation}
Here the symbol $^{0.1}M_r$ denotes the absolute magnitude of the
galaxies in $h=1$ units\footnote{The true absolute magnitude
$^{0.1}M_r^{\true}=\,^{0.1}M_r+5\logh$ where $h=H_0/(100\kmsmpc)$.},
$M_*^{\tofour}=-20.83$ and the parameters $(A,B,C)=(0.85,0.15,-0.04)$.
\footnote{The value of $A$ quoted in T04 is $0.895$. We
believe that this is a typographical error as using $A=0.895$ and
$B=0.15$ incorrectly yields $b(L_*^{\tofour})=1.045 b_*$.}

The total galaxy power spectrum was corrected for the luminosity
dependence of the bias to obtain the galaxy power spectrum for $L_*$
galaxies. The amplitude of the power spectrum of galaxies can be
characterised by the variance of the galaxy density field smoothed on
a scale of $8~\mpch$, $\sigma_{8,g}$. The power spectrum measurements
from T04 yield $\sigma_{8,g}=0.89\pm0.02$ at the effective redshift of
SDSS ($z_*$). This implies that
$b(L_*^{\tofour},z_*)\sigma_8D(z_*)=0.89 \pm 0.02$ where $D(z)$
denotes the growth factor at redshift $z$.

As galaxy bias depends upon redshift it is important to understand the
procedure by which T04 obtain the luminosity dependence of galaxy
bias. The galaxy power spectrum in different luminosity bins obtained
by T04 was fit with a reference matter power spectrum with
cosmological parameters $\Gamma=\Omega_m h=0.213, h=0.72, n_s=1,
\Omega_b/\Omega_m=0.17$ and an amplitude which was allowed to vary
freely. As the galaxies in different luminosity bins are at different
effective redshifts, this implies that the fitting function provided
by T04 (Eq.~\ref{eq:teg04}) should in reality be interpreted as the
ratio $[P^{\g\g}(L,z)/P^{\g\g}(L_*^{\tofour},z_*)]^{1/2}$ where $z$
and $z_*$ are the average redshifts of the bins centered on $L$ and
$L_*$, respectively. This ratio is related to the galaxy bias in the
following manner,
\begin{equation}
\label{eq:teg04act}
\left[\frac{P^{\g\g}(L,z)}{P^{\g\g}(L_*^{T04},z_*)}\right]^{1/2}=\frac{b(L,z)}{b(L_*^{\tofour},z_*)}\frac{D(z)}{D(z_*)}\,.
\end{equation}


\subsection{Small-scale clustering measurements}
\label{sec:diff_dr7}

The measurement of the small scale clustering of galaxies from the
SDSS-DR7 main galaxy sample was presented in \citet[][Z10
hereafter]{Zehavi2010}. Z10 measure the galaxy clustering on scales of
$0.2-30~\mpch$ (projected along the line-of-sight) using volume-limited
samples of galaxies complete above a given luminosity threshold. The
galaxy clustering measurements were modeled analytically using the
halo model. The HOD models predict that the large-scale bias
for galaxies in a luminosity bin is given by
\begin{equation}
b([L_1,L_2]) = \frac{ \int \avn\,n(M,z)\,b(M,z)\,\drm M}
{\int \avn\,n(M,z)\,\drm M } \,,
\label{eq:zeh}
\end{equation}
where $\avn$ is the HOD for the luminosity bin (obtained by
subtracting the HODs of the luminosity threshold samples that bracket
the bin), while $n(M,z)$ and $b(M,z)$ denote the halo mass function
and the halo bias function at the average redshift $z$ of the
luminosity bin sample, respectively \citep{Tinker2008,Tinker2010}. Z10
analyzed their data using the cosmological parameters, $\Omega_m=0.25,
h=0.7, n_s=0.95, \Omega_b=0.045, \sigma_8=0.8$. Using the HOD
parameters given in table 3 of Z10, the large-scale bias of galaxies
can be computed using Eq.~\ref{eq:zeh} given above. This bias should
be interpreted as $b(L,z)(\sigma_8/0.8)$ and will be valid for the
cosmological parameters used by Z10.

Additionally, Z10 also present two other measurements of the galaxy
bias which are not sensitive to the details of their halo occupation
distribution modelling. Z10 use the ratio of the measured projected
galaxy clustering signal in a given luminosity bin ($w_p^{\g\g}$) to
the projected non-linear matter clustering signal ($w_p^{\rm NL}$) at
$r=2.67~\mpch$ to obtain an estimate of the galaxy bias,
$b_{2.67}(L)$. The non-linear matter clustering signal was obtained by
taking a fourier transform of the non-linear matter power spectrum
calibrated by \citet{Smith2003}. Yet another estimate of the galaxy
bias, $b_{\ft}(L)$ was obtained by fitting the ratio $w_p^{\g\g}/w_p^{\rm
NL}$ on scales between $4$ and $30~\mpch$.

Z10 claim that the results obtained from these HOD model-independent
measurements are in agreement with the HOD modeling method. Note that
the HOD modeling results have statistical errorbars that are much
smaller than those obtained from the model independent measurements.
Therefore, the statistical significance of any discrepancy we may find
between the galaxy bias results from T04 and these model independent
measurements is a conservative estimate of the significance of the
discrepancy between the results from T04 and the HOD modeling results
of Z10.

\begin{table}
\begin{center}
\label{tab1}
\caption{Volume limited samples}
\begin{tabular}{ccc}
\hline\hline
  $^{0.1}M_r - 5 \log h$ & [$z_{\rm min}$,$z_{\rm max}$] & [$z_{\rm
  min}$,$z_{\rm max}$] \\
                         & T04 & Z10                \\
\hline
 $(-19.0,-18.0]$ & [0.017, 0.042] & [0.017, 0.042] \\
 $(-20.0,-19.0]$ & [0.027, 0.065] & [0.027, 0.064] \\
 $(-21.0,-20.0]$ & [0.042, 0.103] & [0.042, 0.106] \\
 $(-22.0,-21.0]$ & [0.065, 0.157] & [0.066, 0.159] \\
 $(-23.0,-22.0]$ & [0.104, 0.238] & [0.103, 0.245] \\
\hline\hline
\end{tabular}
\begin{minipage}{\hsize}
  The redshift ranges for the various volume limited samples used by
  Z10 and T04 for their analyses.
\end{minipage}
\end{center}
\end{table}

\begin{figure*} [tc]
\centering
\includegraphics[scale=0.5]{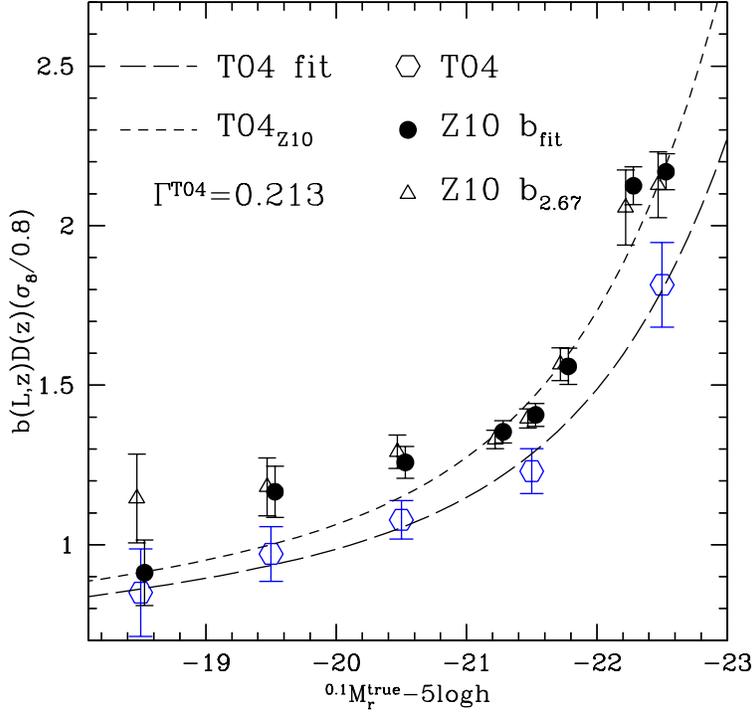} 
\caption{The relation $b(L,z)D(z)(\sigma_8/0.8)$ obtained by different
methods. The open hexagons show the results from large scale power
spectrum measurement of T04 while the long-dashed line is their fit to
these results. The triangles show the ratio between the galaxy
correlation function measured by Z10 and the non-linear matter
correlation function at $2.67~\mpch$. The filled circles show the fit
to this ratio on scales between $2.67$ to $30.0~\mpch$. These two
measurements are model-independent measures of the galaxy bias using
the methods adopted by Z10 but the same cosmological parameters as T04
to allow a fair comparison. The short dashed line shows how Z10
interpret the T04 fit results. 
}
\label{fig1}
\end{figure*}

\begin{figure*} [tc]
\centering
\includegraphics[scale=0.5]{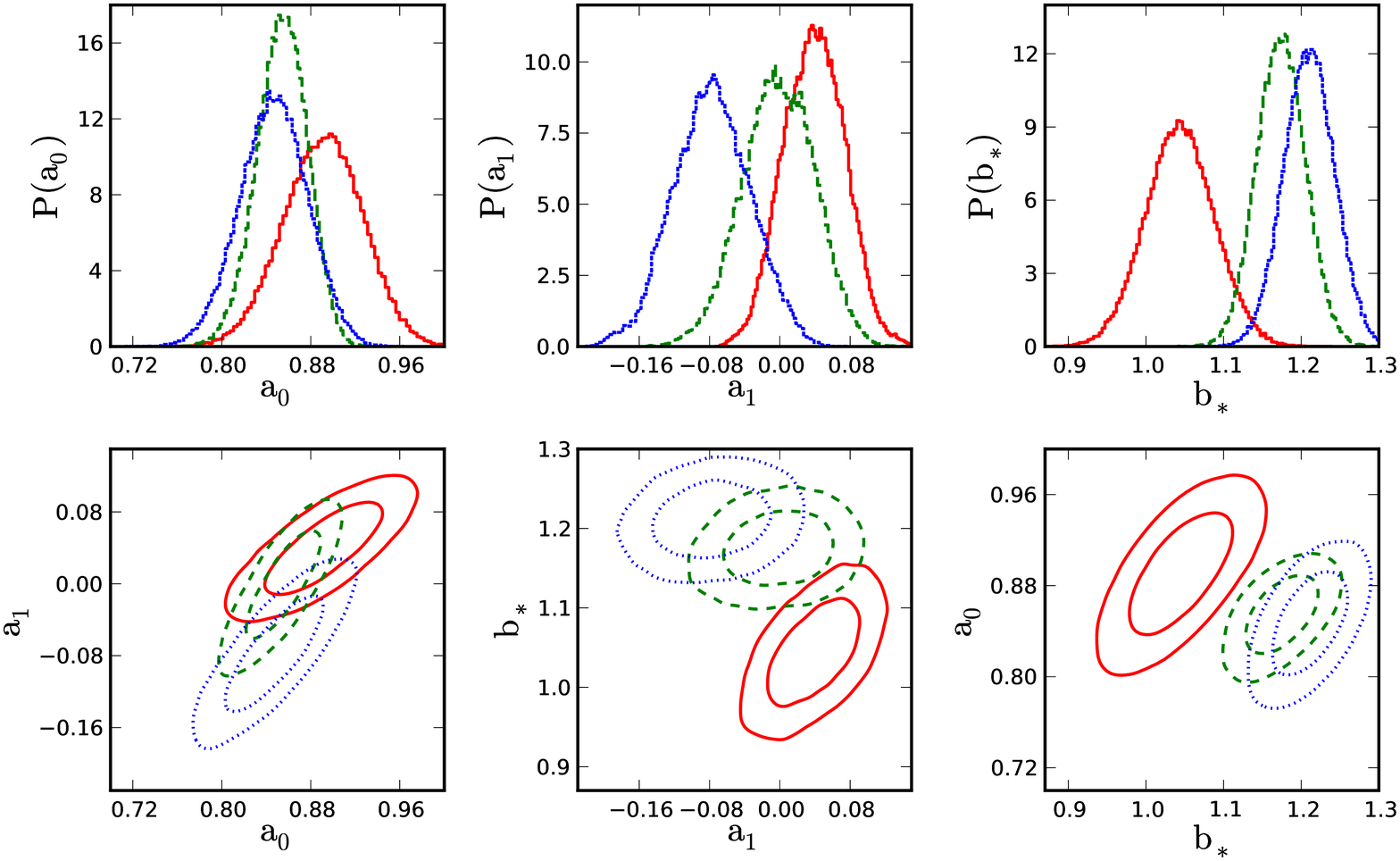} 
\caption{Parametric fits $b(L)=b(L_*^{\rm B03})[a_0+(1-a_0)L/L_*^{\rm
B03}+a_1(M-M_*^{\rm B03})]$ with $M_*^{\rm B03}=-20.44$ to the data
from Fig.~\ref{fig1}. Solid, dashed and dotted lines correspond to
fits to T04, Z10 $b_{\ft}$ and Z10 $b_{2.67}$ data, respectively. The
upper panel shows the probability distributions for the parameters
while the lower panel shows the 68 and 95\% confidence regions in the
parameter space.}
\label{fig2}
\end{figure*}

\subsection{Comparison}
\label{sec:compare}

We now compare the luminosity dependence of the bias, $b(L)$, obtained
by these different methods. We will compare both the shape and the
amplitude of this relation in contrast to the comparison presented by
Z10 in fig. 7 of their paper. There are a couple of issues which we
have to be careful about before we compare the results from the two
different galaxy bias measurements: (i) each of the galaxy bias
measurements is valid at the effective redshift of the luminosity bin
sample under consideration and certain assumptions regarding the
evolution of galaxy clustering need to be made to enable a fair
comparison and (ii) the cosmological parameters used by T04 and Z10
are different which can affect the shape of the galaxy power spectrum
and hence the bias measurements.

The assumption of constant galaxy clustering (CGC) is often used to
interpret and use the galaxy bias results
\citep{Lahav2002,Percival04,Reid2010}. According to CGC, the
clustering of galaxies does not evolve with redshift, i.e.,
$P^{\g\g}(L,z)=P^{\g\g}(L,z_0)$. This implies 
\begin{equation}
b(L,z)=b(L,z_0)\,\frac{D(z_0)}{D(z)}\,.
\end{equation} 
We point out that it is only under the CGC assumption, that the ratio
on the left hand side of Eq.~\ref{eq:teg04act} can be equated to
$b(L)/b(L_*^\tofour)$, such that both biases are measured at the same
redshift. The results from T04 have been presented (and often used) in
this manner without stating the underlying CGC assumption \citep[see
e.g.,][]{Hand2011}.

However, we can proceed with our comparison without making
the (perhaps questionable) CGC assumption. This is because each of the
luminosity bin samples from Z10 is at the same effective redshift as
that of the corresponding luminosity bin sample from T04. This can be
seen from the redshift ranges used by T04 and Z10 for constructing
their volume limited samples (see Table~\ref{tab1}). The differences in
the redshift ranges which arise because Z10 use all galaxies with
r-band apparent magnitudes $14.5<m_r<17.6$ while T04 use galaxies with
$14.5<m_r<17.7$, are very small and can be safely ignored. This allows
us to compare the quantity $b(L,z)D(z)(\sigma_8/0.8)$ obtained by
these authors without assuming a specific form for the redshift
evolution of galaxy clustering.

To address the second issue, we use the same cosmological parameters
as were used by T04 in their analysis and carry out the measurement of
$b_{2.67}$ and $b_{\ft}$ from the galaxy-galaxy clustering
measurements and the covariance matrices obtained by Z10 \footnote{The
data was kindly provided in an electronic format by I. Zehavi. The
measurement of the galaxy clustering for the luminosity bin
$[-21,-20]$ was carried out for two different samples by Z10, one that
includes the Sloan Great Wall (SGW), the other one which excludes it.
We use the galaxy clustering measurements obtained from the sample
that includes the SGW as T04 carried out their measurements on such a
sample as well.}. To calculate $b_{2.67}$ we divide the projected galaxy
clustering measurement with the projected matter clustering
signal computed from the non-linear matter power spectrum of
\citet{Smith2003} at redshift $z=0$. To calculate the galaxy bias
using the $b_{\ft}$ method, we find a scale-independent
single-parameter fit to the ratio of the projected galaxy clustering
to the projected non-linear matter clustering on scales between $4$ and $30
~\mpch$ ($b_{\ft}[L]$). We account for the covariance of the data
points. The methods used by us are identical to those used by Z10 with
the only difference that we use the cosmological parameters from T04
and the $z=0$ matter correlation function. This procedure yields
us $b(L,z)D(z)(\sigma_8/0.8)$. We do not carry out the full HOD
modeling but the results from such modeling should agree fairly well
with the $b_{\ft}$ and $b_{2.67}$ measurements as shown by Z10.


The results from T04 can be recast as
\begin{eqnarray}
b(L,z)D(z)\frac{\sigma_8}{0.8}&=&\left[\frac{b(L,z)D(z)}{b(L_*^{\tofour},z_*)D(z_*)}\right]\frac{b(L_*^{\tofour},z_*)D(z_*)\sigma_8}{0.8}\nonumber\\&=&\left[\frac{b(L,z)D(z)}{b(L_*^{\tofour},z_*)D(z_*)}\right]\frac{0.89\pm0.02}{0.8}\,.
\end{eqnarray}
Here, for the second equality we have used $\sigma_{8,g}=0.89\pm0.02$
as obtained by T04. The actual measurements of the quantity in the
square brackets obtained from the galaxy power spectra by T04 were
reported in table 1 of \citet{Seljak2005}. We use these measurements
and the above equation to obtain the blue hexagons with errorbars
shown in Fig.~\ref{fig1}.  The fitting function (Eq.~\ref{eq:teg04})
given by T04 is shown with a long-dashed line.

The results based on the model independent methods of Z10 are shown as
triangles with errorbars ($b_{2.67}$) and as filled circles with
errorbars ($b_{\ft}$), respectively. In figure 7 of their paper, Z10 also
compare their results with T04. While plotting the galaxy
bias-luminosity relation from T04, they use Eq.~\ref{eq:teg04} with
the parameters $(A,B,C)=(0.85,0.15,-0.04)$ but use $M_*=-20.5$ instead
of $-20.83$, originally used by T04. They also use the galaxy bias
obtained from their analysis to get $b(L)$ from $b/b_*$ instead of
using the $b_*$ obtained by T04. The galaxy bias from T04 as
interpreted by Z10 in this manner is shown with a dashed line. They
claim agreement between their results and T04 based upon this line.


However, it is clear that their comparison tests only for the
consistency between $b/b_*$ as opposed to the comparison we present
where we check both the normalization and the shape of the galaxy
bias-luminosity relation. It can be easily seen from Fig.~\ref{fig1}
that the luminosity dependence of the large-scale bias $b(L)$ obtained
from the large-scale power spectrum method (T04) is systematically
lower than that obtained from the small scale clustering methods
(Z10). To assess the systematics quantitatively, we show the results
of fitting a parametric form to the galaxy bias obtained by different
methods in Fig.~\ref{fig2}. We consistently use the magnitude of $L_*$
galaxies to be equal to $M_*=-20.44$ as reported by
\citet{Blanton2003} when analysing the galaxy bias data from T04 or
Z10 \footnote{See the discussion section to understand the source of
the different values of the magnitude of $L_*$ galaxies used in the
literature.}. Although Fig.~\ref{fig1} also shows the galaxy bias
data obtained from the small scale clustering at few
intermediate luminosity bins, we only use the data from the luminosity bins
listed in Table~\ref{tab1} while carrying out the fit.  The value of $b_*=b(M_*^{\rm B03})$ obtained from the
T04 data is $1.04\pm0.04$ while that obtained from the $b_{\ft}$
($b_{2.67}$) data from Z10 is $1.17\pm0.03$ ($1.21\pm0.03$), in
significant tension with each other.  The other parameters that
describe the shape of $b(L)$ relation are also at best only marginally
consistent as can be seen from the 68 and 95\% confidence contours
shown in the lower panels.  We point out that these systematic
differences were not noticed earlier as the T04 results have often
been interpreted as being valid for galaxies with $^{0.1}M_r=-20.5$
and the normalization of the galaxy bias from T04 was not used for the
comparison \citep[see][Z10]{Zehavi2005}.

Some of the differences at the faint end could be attributed to cosmic
variance and differences in the data releases \citep{Meng2010}.
However, the systematic differences are as large as $\sim15\%$ for
galaxies with magnitude $^{0.1}M_r=-20.5$ and are seen to increase to
$\sim20\%$ as we consider brighter galaxies.  The small-scale
clustering results based on SDSS-DR2 by \citet[][hereafter
Z05]{Zehavi2005} show similar differences at the bright end when
compared to T04. A marginal hint of the difference in galaxy bias at
the bright end between T04 and Z05 was previously noticed by
\citet{Swanson2008}. These authors used a counts-in-cells analysis to
obtain a relative galaxy bias ($b/b_*$) using galaxies from SDSS.
Their galaxy bias-luminosity relation agrees well with results from
T04. However, the discrepancy with Z05 was not emphasized much by
\citet{Swanson2008} as the galaxy bias measurements from Z05 were
inferred from the clustering of galaxies at $2.67~\mpch$ where a
possible residual scale dependence of the bias cannot be ruled out.
Note that unlike the $b_{2.67}$ measurements, the $b_{\ft}$
measurements from Z10 should be more robust, and the HOD modelling
based galaxy bias should in principle have accounted for the scale
dependence of bias if any. Therefore, the discrepancy can no longer
be swept under the rug based upon the scale dependence of bias. We
also note that \citet{Swanson2008} did not explicitly check agreement
between the normalization $b_*$ of the galaxy bias between T04 and
Z05, which can reduce the significance of the discrepancy.

\begin{figure} [tc]
\centering
\includegraphics{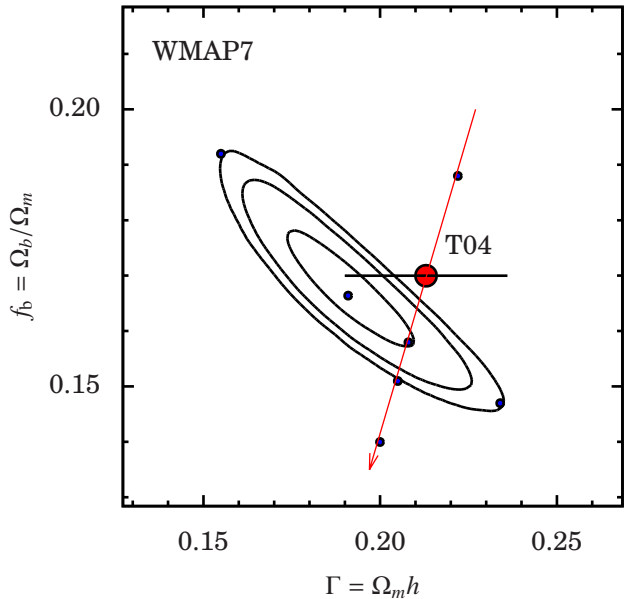} 
\caption{
    Confidence contours in the $\Gamma-f_{\rm b}$ plane obtained from
    the Monte Carlo Markov chains of \citet{Komatsu2011} who analysed
    the 7 year CMB data from WMAP. The red circle with errorbars shows
    the constraint on $\Gamma$ obtained by T04 under the assumption of
    a $\delta$-function prior on $f_b$. The red arrow shows the
    general direction of the discrepancy between $\Gamma$ and $f_{\rm
    b}$ if both parameters are allowed to be free when analysing the
    power spectrum data from T04.  The small blue circles show the
    cosmological parameters we used to test if the discrepancy between
    the galaxy bias measurements depends upon the cosmological
    parameters used to carry out the analysis.
}
\label{fig:cosmo}
\end{figure}

\begin{figure*} [h]
\centering
\includegraphics[scale=0.75]{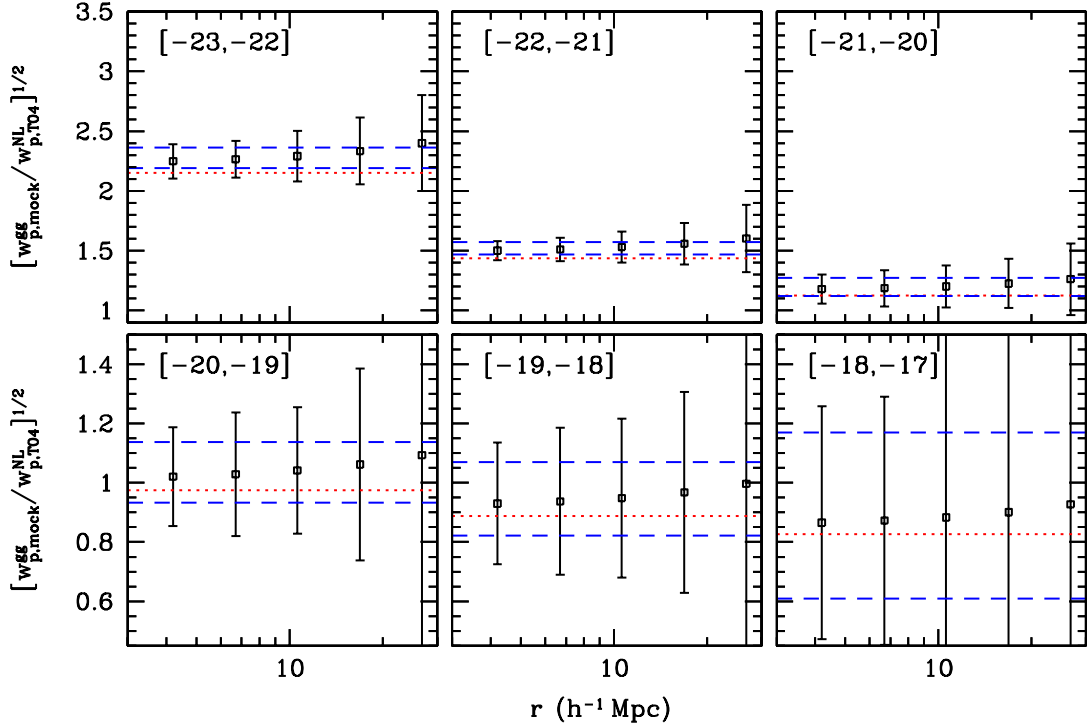} 
\caption{
    The points with errorbars show the ratio of the projected galaxy
    clustering in the fiducial mock model (cosmological parameters
    consistent with WMAP) to the projected non-linear matter
    clustering assuming the cosmological parameters from T04.
    Different panels correspond to different luminosity bins, the
    brightest one is in the top left corner while the faintest one is
    in the bottom right corner. The red dotted line shows the "true"
    galaxy bias in our mock model. The blue dashed lines show the 68
    percent confidence interval derived from fitting the data (see
    text for details).
}
\label{fig:wpr}
\end{figure*}

\begin{figure*} [h]
\centering
\includegraphics[scale=0.75]{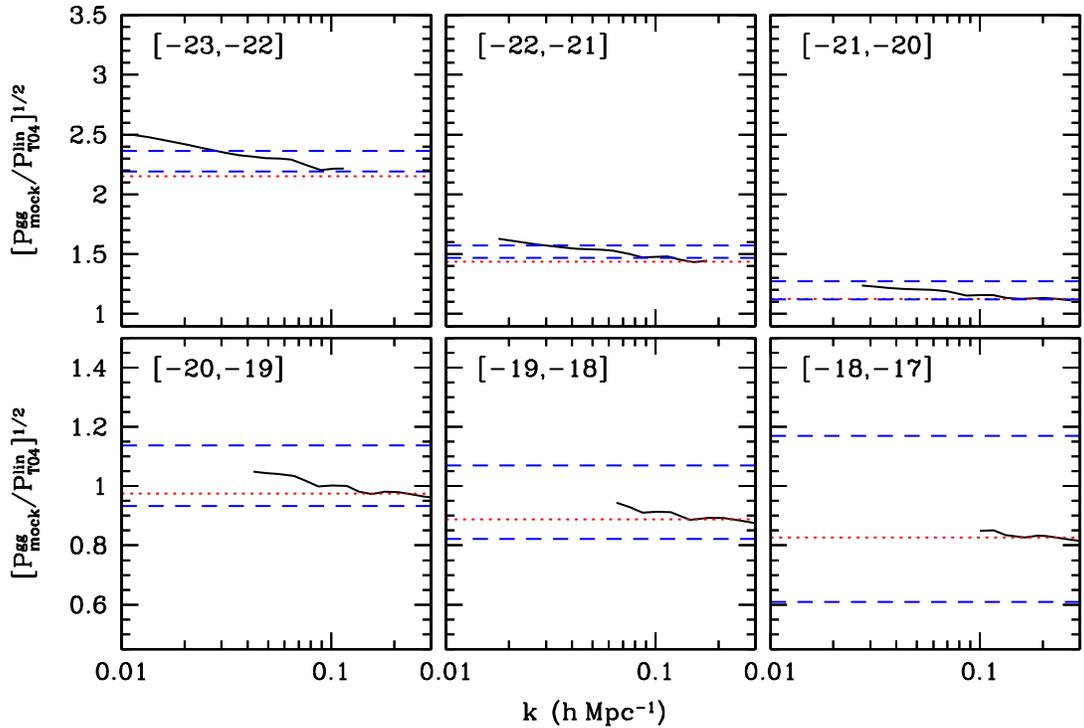} 
\caption{
    Solid lines show the ratio of the galaxy-galaxy linear power
    spectrum in the fiducial mock model (cosmological parameters
    consistent with WMAP7) to the linear power spectrum assuming the
    cosmological parameters from T04. Different panels correspond to
    different luminosity bins. The red dotted line shows the "true"
    galaxy bias in our mock model. The blue dashed lines show the 68
    percent confidence interval derived from fitting the small scale
    clustering data from Figure~\ref{fig:wpr}.
}
\label{fig:pklin}
\end{figure*}

Finally, we investigate the possibility that the discrepancy between
the galaxy bias measurements from the two methods is a result of using
the cosmological parameter set from T04 instead of the "concordance"
parameters that are currently favored by the cosmological datasets.
Figure~\ref{fig:cosmo} shows the 68, 95 and 99 percent confidence
contours obtained by using the Monte-Carlo Markov chains of
\citet{Komatsu2011} from the analysis of the 7 year cosmic microwave
background data from WMAP. We show the results in the $\Gamma-f_b$
plane where $\Gamma=\Omega_m h$ is the shape parameter, and
$f_b=\Omega_b/\Omega_m$ is the universal baryon fraction. These two
parameters primarily determine the shape of the linear power spectrum.
The big red circle shows the cosmological parameters that we adopted
from T04 to analyse the data from Z10. T04 had adopted a
$\delta$-function prior of $f_{\rm b}=0.17$ while analysing their
data. The errorbar on the big red circle in the x-direction shows the
uncertainty quoted on the parameter $\Gamma$ by T04. The red arrow
shows the general direction of the degeneracy that can be obtained
based on the measurements of the shape of the power spectrum when both
$\Gamma$ and $f_{\rm b}$ are used as free parameters of the model (see
Figure 38 from T04).

To test the sensitivity of the discrepancy to a change in cosmological
parameters, the following test can be carried out. Both the galaxy
clustering data from Z10 and the power spectrum data from T04 can be
analysed in the framework of the maximum likelihood cosmological
parameters from WMAP7 and the resultant galaxy biases can be compared
with each other. Unfortunately, the data for the galaxy power spectrum
of the six volume limited samples from T04 could not be obtained from
the corresponding author. In principle, the measurements from Figure
28 of T04 can be obtained with the use of applications such as Dexter
\citep{Demleitner2001}. However, the errorbars between data points
are highly correlated and the data obtained in this manner without the
full covariance matrix is therefore of very limited use.

As the above straightforward test could not be carried out, we
performed the following test instead. We use a fiducial cosmological
model and generate a mock data set for the projected galaxy clustering
on scales from $3~\mpch$ to $30~\mpch$ and the galaxy-galaxy linear
power spectrum signal on the same scales used by T04. We then analyze
these data assuming the cosmological parameters of T04 and check
whether the galaxy bias measurement agrees between the two methods.
Our tests, which we describe below in detail, show that the
discrepancy could not have been a result of the assumption of
cosmological parameters which differ from the true underlying
parameters. Readers not interested in the details of these tests may
skip the rest of this section.

As a fiducial model, we used the cosmological parameters which give
the maximum likelihood for the WMAP7 data, $\Omega_m=0.271,
\Omega_b=0.0451$, $n_s=0.966$, $h=0.703$, $\sigma_8=0.809$. We
calculated the value of the projected non-linear matter clustering at the
radii where Z10 measure the projected galaxy clustering. We then used
a mock galaxy bias-luminosity relation with the same parameterization
as Equation~\ref{eq:teg04}. We use the parameters $b_*=1.125$,
$A=0.85$, $B=0.15$ and $C=-0.04$, and assume the magnitude of $L_*$
galaxies to be $-20.44$. We then used this bias to calculate the
projected galaxy clustering expected in this model.  We assigned each
clustering data point a relative error which is equal to that obtained
by Z10 in their data (we used only the diagonal part of the covariance
matrix for this test). 

This data was then analyzed by using the
cosmological parameter set from T04. The ratio of our mock projected
galaxy clustering data to that of the projected non-linear matter clustering
(using the cosmology from T04) is shown in Figure~\ref{fig:wpr}. The
errorbars show the relative errors on the measurements. The red
dotted line shows the input (true) galaxy bias. To calculate the
uncertainty with which the galaxy bias would be measured from data of
this quality, we generated 10000 Monte Carlo datasets for each of the
luminosity bins using the errorbars shown in Figure~\ref{fig:wpr}. We
fit each of the luminosity bin from every dataset using a single
parameter, $b$ for the bias. For every luminosity bin, this gives us a
probability distribution for the value of bias, whose 16th and 84th
percentile (corresponding to the 1 $\sigma$ region) are shown using
the blue dashed bands. 

Next, we computed the linear power spectrum of galaxies in different
luminosity bins using our fiducial cosmological model and our mock
bias model at the values of $k$ (which are different for different
luminosity bins) where T04 measure the galaxy power spectrum. We
divided these galaxy power spectra by the linear matter power spectrum
using the cosmological parameters of T04.  These are shown as solid
lines in Figure~\ref{fig:pklin}. The blue dashed lines are the result
of fitting the galaxy bias from the small scale clustering method and
are the same as those seen in Figure~\ref{fig:wpr}. If there was a
systematic difference between the measurement of galaxy bias between
the power spectrum method and the clustering method just due to the
assumption of a cosmology different from the fiducial cosmological
model, the solid line in Figure~\ref{fig:pklin} should have fallen
systematically outside the confidence region shown by the blue dashed
lines.

We also tried other cosmological parameters to create the mock dataset
(these are marked with the blue small circles in
Figure~\ref{fig:cosmo}) and found no appreciable difference from the
conclusion stated above. Whenever parameters that lie along the the
red arrow are chosen to generate the mock data and the T04 set of
parameters is used to analyse the data, the inferred galaxy bias from
the power spectrum measurements does not show an appreciable scale
dependence, i.e. the solid line in Figure~\ref{fig:pklin} appears
flat. The galaxy bias measured from the power spectrum lies well
within the uncertainty on the galaxy bias inferred from the small
scale clustering measurements. The largest departure from scale
dependence is observed when using the cosmological parameters
corresponding to the blue filled circle at the left hand corner of the
WMAP7 confidence contours as the fiducial set.  However, even in that
case, the bias measurements from the two methods appear to be in fair
agreement with each other. This test shows that the discrepancy
between the galaxy bias from the small scale clustering measurements
and the galaxy power spectrum measurements is not due to our
assumption of the cosmological parameters from T04 to analyse the
data. 


\section{Summary and Discussion}
\label{sec:discuss}

We have compiled measurements of the luminosity dependence of the
galaxy bias obtained using various methods and compared the shape and
the normalization of the galaxy bias. We find that the bias of $L_*$
galaxies with magnitude $^{0.1}M_r=-20.44$ is at best constrained to
an accuracy of order 10-15\% (considering both statistical and
systematic errors). We find that the galaxy bias obtained from
large-scale power spectrum methods is systematically lower than that
obtained from the small-scale clustering measurements. The discrepancy
increases as a function of galaxy luminosity. We have also shown that
this systematic discrepancy is not a result of assuming cosmological
parameters which are different from the concordance cosmological
parameters. We briefly comment on the plausible origins of the
discrepancy and its consequences.

\subsection{Typographical errors}
We first examine whether typographical errors in any of the
manuscripts could have caused this discrepancy. The number which has
the potential to raise a few eyebrows is the value of $M_*$ used by
T04. The luminosity function has been measured with a great accuracy
using SDSS data by \citet{Blanton2003} who obtain $M_*=-20.44\pm0.01$,
instead of $-20.83$ used by T04.  We carried out the following
exercise to verify that the $M_*$ to be used in Eq.~\ref{eq:teg04}
from T04 is indeed $-20.83$. Using the values of $b/b_*$ obtained by
T04, our own fitting routines and assuming $M_*=-20.83$, we obtain the
values of parameters $(A,B,C)$ in Eq.\ref{eq:teg04} to be
$(0.856,0.144,-0.038)$ consistent with T04. Instead, if we use
$M_*=-20.44$, we obtain $(A,B,C)=(0.928,0.072,-0.068)$ and a fit which
is significantly worse than the former. This rules out the possibility
of typographical error by T04 while quoting the value $M_*$.

The value of $M_*=-20.83$ used by T04 \citep[and also later
by][]{Swanson2008} appears to be taken from
\citet{Blanton2001} where the luminosity function of SDSS galaxies was
measured from early commissioning data (M. Blanton 2011, private
communication). This value of $M_*$ was later rectified in
\citet{Blanton2003} by including a model for the passive evolution of
galaxies and using a fitting function more flexible than the Schechter
function. It appears that T04 failed to incorporate this change in
their analysis.  However, this does not invalidate their power
spectrum measurements. The power spectrum presented in T04 should be
interpreted as that of galaxies with $^{0.1}M_r=-20.83$ as we have
done in this paper, instead of that of $L_*$ galaxies.

\subsection{Systematic issues: Power spectrum method}
The large scale power spectrum measurements from T04 may also have
some systematic uncertainties. To calculate the three dimensional
power spectrum of galaxies, T04 had to remove the small scale redshift
space distortions (finger-of-god effect, FOG hereafter) induced by the
peculiar velocities of the galaxies. Their method involved using the
friends-of-friends algorithm \citep{Huchra1982} in redshift
space to identify overdense regions (the overdensity $\delta_c$ is a
free parameter of the algorithm, T04 used $\delta_c=200$ as the
fiducial value) in the galaxy distribution. The FOG removal algorithm
then measures the dispersion of galaxy positions about the center of
the identified overdense region in both the radial and transverse
directions and then compresses the positions of galaxies radially
until the dispersions are equal in both the transverse and radial
directions. T04 used an anisotropic metric to measure distances
between galaxies to account for the anisotropic stretching of galaxies
in the line-of-sight direction. Two galaxies were
marked as friends of each other if their separations satisfy
\begin{eqnarray}
\left[ \left(\frac{r_{||}}{10}\right)^2 + r_{\perp}^2 \right]^{1/2} &\leq&
\left[ \frac{4}{3}\pi(1+\delta_c)\bar{n} + r_{\rm \perp max}^{-3}
\right]^{-1/3} \\
&\leq& \left[ \frac{4}{3}\pi(1+\delta_c) +\left(\frac{r_{\rm \perp
max}}{\bar{l}}\right)^{-3}\right]^{-1/3} \bar{l}
\end{eqnarray}
where, $\bar{n}$ is the mean number density of galaxies,
$\bar{l}=\bar{n}^{-1/3}$ is the mean galaxy separation, $r_\perp$
is the projected distance, $r_{||}$ is the distance along the
line-of-sight and $r_{\rm \perp max}$ was set to be equal to $5~\mpch$
by T04. This implies that the value of the linking length in units of
the mean galaxy separation when $\delta_c=200$ is $\approx 0.106$
(ignoring the $r_{\rm \perp max}$ term which is negligible compared to
the $\delta_c$ term).  For such a small value of linking length the
overdensity of structures identified by FOF should be $\sim 2000$,
i.e. much larger than $200$ \citep{More2011} which T04 aim to find and
this can be a possible cause of systematics. \footnote{A C++ code to
calculate the overdensity of FOF haloes for any given cosmology and
linking length parameter is available at
\href{http://kicp.uchicago.edu/~surhud/research/odcode.tgz}{http://kicp.uchicago.edu/$\sim$surhud/research/odcode.tgz}.}

The measurements of the galaxy bias by \citet{Swanson2008} based on a
counts-in-cells analysis, which agree with T04, also applied the same
algorithm for FOG removal.  It must be noted however that T04 included
a check for systematics by changing the value of $\delta_c$ to be
lower than their fiducial value which causes the linking length to be
larger than $0.106$ and did not find large systematic effects on the
measured galaxy power spectrum compared to the statistical errorbars
(see Figures $20$ and $21$ in T04). 

\subsection{Systematic issues: Small scale clustering}
Modeling the small scale clustering of galaxies is inherently
difficult as it requires knowledge of the highly uncertain physics of
galaxy formation. The halo model conveniently by-passes this issue by
using a statistical description of how galaxies populate halos as
opposed to a full fledged model for galaxy formation physics.
However, such a modeling has its own issues. Its unclear how
differences in the assumed parametric forms for the HOD affect the
modeling results. On small scales, pairs of galaxies in the same halo
dominate the clustering signal. An accurate knowledge of the number
density distribution of galaxies in a single halo and the satellite
fraction (fraction of galaxies of a given luminosity which are
satellites) are required to model this signal \citep[e.g.,
see][]{Mandelbaum2006, Yang2008, More2009}. The occupation number of
satellites in halos of a given mass is often assumed to be distributed
in a Poisson manner, which is perhaps a questionable assumption
\citep[e.g., see][]{Wetzel2010}. The transition region where the
clustering of galaxies is equally composed of pairs from the same halo
and those from separate halos has also been notoriously difficult to
model. An accurate knowledge of the scale dependence of the bias and
an accurate treatment of halo exclusion is central to model the
clustering in this region \citep{Tinker2005, Smith2011}. Given all
these complications, it is important to test for consistency of the
small scale clustering results with the independent measurement from
the large-scale power spectrum of galaxies, especially as precision
constraints on cosmological parameters are being obtained from the
modeling of small-scale clustering \citep[see e.g.,][]{Tinker2011}.

Another potential systematic effect concerns the range of integration
used along the line-of-sight, $\pi_{\rm max}$, to project the
3-dimensional correlation function to obtain $w_p(r_p)$. In Figure $1$
of \citet{Norberg2009}, the authors show that the projection of the
correlation function by integrating in redshift space upto $\pi_{\rm
max}=64~\mpch$ gives results that are systematically larger compared
to results when the integration is carried out in real space. The difference
between these results is as large as $10\%$ on scales of
$r_p=10~\mpch$ with differences which increase with scale and grow
as large as $30\%$ at scales of $30~\mpch$. The projection is
supposed to get rid of the redshift distortions but only when the
integration is done along the entire line-of-sight. However, when the
integration is done only out to a certain maximum value ($\pi_{\rm
max}=60~\mpch$ is a commonly used value), the large scale redshift
distortions (known as the Kaiser effect) may still persist. This can
cause an overestimate of the correlation function corresponding to an
overestimate of the galaxy bias. The exact overestimate will depend
upon the quantity
\begin{equation}
\beta = \frac{1}{b}\frac{\drm \ln
D(z)}{ \drm \ln (1+z)^{-1} } \,.
\end{equation}
Our preliminary estimates of the effect indicate that it can reduce
the discrepancy between the two measures of the galaxy bias but not
entirely eliminate it. The galaxy bias measurements from Z10 when the
Kaiser effect is included should be smaller than those shown in
Figure~\ref{fig1} by roughly 1 $\sigma$. We intend to pursue the issue
of estimating the magnitude of the Kaiser effect on the projected
clustering measurements, its dependence on $\pi_{\rm max}$ and
comparisons with N-body simulations in future work.

\subsection{Possible implications}
The luminosity dependence of the galaxy bias obtained by T04 was used
in conjunction with HOD modeling of the galaxy-galaxy lensing results
by \citet{Seljak2005} to obtain constraints in the $b_*-\sigma_8$
plane. It is unclear how the likelihood contours obtained by these
authors will shift in the $b_*-\sigma_8$ plane if they were to use the
small scale clustering results instead. Future analyses which combine
galaxy-galaxy lensing and clustering data should be able to shed more
light on this issue. The luminosity dependence of the galaxy bias has
also been used to obtain an effective halo mass for galaxies of a
given luminosity \citep[e.g., see][]{Hand2011}. One should be aware
that the systematics highlighted in this paper will lead to large
uncertainties in the halo masses especially at the bright end.

The LRG power spectrum measurements \citep{Reid2010} use the $b(L)$
obtained by T04 to correct for the red-tilt of the galaxy power
spectrum.  Systematic uncertainties in the shape of $b(L)$ shown in
this paper can cause a residual scale-dependence in the galaxy power
spectrum modifying its shape and systematically bias the estimates of
the cosmological parameters derived from it. A consistent
understanding of galaxy bias and the systematics associated with each
of the methods is therefore crucial to obtain precision constraints on
cosmological parameters from the galaxy distribution.


\section*{Acknowledgments}

The author is supported by the KICP through the NSF grant PHY-0551142
and an endowment from the Kavli Foundation. The author thanks the
anonymous referee for useful comments and suggestions which helped
improve the content of the paper. The author also acknowledges useful
discussions/electronic correspondences with Michael Blanton, Nick
Gnedin, Andrey Kravtsov, Will Percival, Neelima Sehgal, Uros Seljak,
Max Tegmark, Frank van den Bosch, Idit Zehavi and Zheng Zheng. The
author also thanks Nick Gnedin, Neelima Sehgal and Frank van den Bosch for
a critical reading of draft versions of the manuscript.



\end{document}